\documentclass{aa}
\usepackage{graphics}
\begin{document}
 
\thesaurus{03(11.05.1;11.09.1;09.16.1)}
\title{Spectra of planetary nebulae in NGC~5128 (Centaurus-A)}
\titlerunning{Spectra of PN in NGC~5128}
\author{J. R. Walsh\inst{1}
\and
N. A. Walton\inst{2}
\and 
G. H. Jacoby\inst{3}
\and
R. F. Peletier\inst{4}
}
\authorrunning{Walsh et al.}
\offprints{J. R. Walsh}
\institute{
Space Telescope European Co-ordinating Facility,
European Southern Observatory,
Karl-Schwarzschild Strasse 2, 
D85748 Garching bei M\"{u}nchen,
Germany.
E-mail: jwalsh@eso.org
\and
Royal Greenwich Observatory, 
Apartado 321, 
Santa Cruz de La Palma, 
38780 Tenerife,
Spain. 
E-mail: naw@ing.iac.es
\and
Kitt Peak National Observatory,
National Optical Astronomy Observatory,
Tucson
AZ 85726
U. S. A.
E-mail: gjacoby@noao.edu
\thanks{Visiting Astronomer,
Steward Observatory,
University of Arizona,
933 N. Cherry Ave,
Tucson,
AZ 85721
U. S. A.}
\and
Dept. of Physics,
University of Durham,
South Road,
Durham 
DH1 3LE
England.
Email: R.F.Peletier@dur.ac.uk
}

\date{ }
\maketitle

\begin{abstract}
Low dispersion spectra have been obtained of five planetary nebulae
in the elliptical galaxy NGC~5128 (Centaurus-A) from the catalogue
of Hui et al. (\cite{huib}). The planetary nebulae (PN) cover a range 
of galactocentric radius from 7.9 to 17.7$'$ (8 to 18Kpc). The spectra 
display typical emission 
lines of H~I, He~I, He~II, [O~III], [N~II] and [S~II] and appear very 
similar to high excitation planetary nebulae in the Galaxy. This implies
that, from a stellar evolution viewpoint, there should be no peculiar 
effects introduced by considering the bright cut-off of the PN 
luminosity function for distance estimation. In particular
the brightest PN detected in NGC~5128 is not spectroscopically
unusual. One of the PN shows relatively 
strong He~II and [N~II] lines and the
derived N/O ratio indicates that it may be a Type I nebula, considered 
to arise from a high mass progenitor star. Determinations of 
the oxygen abundance of the five PN shows a mean value 0.5 dex below 
solar. Given that NGC~5128 is an 
elliptical galaxy with a presumably metal rich stellar content, the
low metallicities of the PN are unexpected, although a similar
situation has been observed in the bulge of M~31. 
\end{abstract}

\section{Introduction}
The presence of
abundance gradients in spiral galaxies is well established from
observations of their H\,II regions (e.g. Pagel \& Edmunds \cite{pe81}).
From emission line strengths, abundances
of He, N, O, Ne \& S can be measured, and typically O/H
decreases with galactocentric distance (e.g. Shaver et al. \cite{shav};
Walsh \& Roy \cite{wr89}). The
situation with regard to early--type galaxies is however more
complex. There is little interstellar medium 
and hence no well--dispersed H\,II regions; resort must be made
to the line of sight integrated properties of the starlight.
Both sets of abundance indicators present their own
advantages and disadvantages: emission line regions can be
subject to local enrichments making them atypical of the
general interstellar medium; stellar indicators, whilst
providing line of sight abundances, are subject to
contributions from stars all over the HR diagram, having a
range in metallicity and age. In addition even a 
contemporaneous stellar population can appear to
possess an abundance spread of light elements such as C, N 
and O as found for globular cluster giants (Kraft \cite{kraf}).

 Stellar abundances are measured from the strength of absorption 
lines, such as  Mg~I and molecular bands
such as CN and TiO (Gorgas et al. \cite{gorg}, Davies et al. 
\cite{dav} or colours (Visvanathan \& Sandage \cite{visa} 
and Peletier et al. \cite{pel}). 
Synthesis techniques (Tinsley \cite{tins}, Faber \cite{fab} and
O'Connell \cite{oco}) are required to provide
abundance determinations.  Abundances of various elements, such as
Fe, Mg, Ca, Na etc can be obtained (see e.g. Vazdekis et al. 
\cite{vaz}). Their accuracy is 
limited by the fact that only strong lines can be used, because of
the considerable velocity broadening in the galaxies.
Stellar colour and absorption line variations across
the faces of early--type galaxies are generally observed and
interpreted by an outwardly--decreasing metallicity (eg.
Davies et al. \cite{dav}, Bica \cite{bic}) although 
significant gradients in age are also sometimes claimed (Worthey
et al. \cite{wor}, Trager et al. \cite{trag}).
Globular clusters cannot be used to determine the stellar population
distribution of ellipticals, since their colours and line strengths
are not representative of the main stellar component. For
example Bridges et al. (\cite{brig}) found that the average metallicity
for the globular clusters in M~104 is [Fe/H]=$-$0.7, about a 
factor 6 - 10 lower than for the stars in the central region 
(Vazdekis et al. \cite{vaz}).

Studies pioneeered by Jacoby, Ciardullo, Ford and
co-workers have shown that in early-type systems, planetary
nebulae (PN) can be useful both as distance indicators 
(e.g. Jacoby \& Ciardullo \cite{jac92}) and as
probes of the galaxy kinematics (e.g. Hui et al. \cite{huic}). 
PN, easily detected from their
strong emission line spectra, can also be used as abundance
tracers in a way comparable to H\,II regions, since some
elements, O in particular, but also Ne and S, are not generally
affected by the nucleosynthetic processing in most PN
progenitor central stars. Measuring abundances of PN
provides a unique way to determine the abundance spread of the
old stellar population in distant ellipticals, as predicted from star
formation theories (Arimoto \& Yoshii \cite{ari})
and population synthesis (Bica \cite{bic}). Direct measurement of
the stellar abundance spread from the ground is limited to local group
galaxies such as M31 and M32 where single stars can feasibly be
resolved.

NGC~5128 (Centaurus-A) is the closest giant elliptical
($\sim$3.5 Mpc, Hui et al. \cite{huia}; morphological type S0,pec
Sandage \& Tammann \cite{sata}), and has a
projected size on the sky of over 1$^{\circ}$, making it ideal
for spatial studies. 
HII regions, star formation and interstellar matter are found in the 
inner regions, possibly arising from a merger with a more metal poor
galaxy (see for example Quillen et al. \cite{quil}), so cannot
provide reliable abundance
diagnostics for the old stellar system. 784 PN have been
detected in an area 20$\times$10kpc (Hui et al. \cite{huib}) 
and the luminosity function within the brightest 1.5
magnitudes was used to determine the distance. The radial velocities
of 433 of these PN have been measured (from the brightest line,
that of [O~III]5007\AA) in order to study the dynamics of the halo
(Hui et al. \cite{huic}). Using the PN as test particles, the 
gravitational potential of the galaxy can be studied: Cen-A
was found to have a tri-axial potential with the galaxy minor axis offset 
from the rotation axis by 40$^\circ$. Measuring the rotation  
and velocity dispersion of the PN system, it was shown that M/L increases
with radius suggesting that dark matter was present in the galaxy
halo (Hui et al. \cite{huic}). This is also in agreement with globular cluster
velocity measurements (Hui et al. \cite{huic} for NGC~5128 and
Bridges et al. \cite{brig} for M~104) 

  So far no spectroscopy of these PN has been obtained and their 
relation to the PN population in the Milky Way, which may have 
a different metallicity and certainly a different star formation 
history to Cen-A, 
is not known. In particular it is not known if the brightest PN observed 
are exceptional (perhaps of Type I) and what is the effect of line of sight 
extinction on the PN luminosity function. In Cen-A, it is known that there
is a jet and extended emission line regions along the jet (e.g.
Morganti et al. \cite{morg}) as well as H\,II regions in the vicinity of the 
dust lane, all of 
which could be included in the PN census at some level. Spectroscopy can 
therefore be seen as important both in terms of the PN population and its
use as a distance indicator and in terms of probing the chemical history of 
the host galaxy. The long-term goal is to study abundance gradients and the
spread in abundances at a given radius, from 
large numbers of PN observed with multi-object techniques. 

In this paper the results of deep spectroscopic integrations with a 3.6m
telescope of a few selected
PN in NGC~5128 are presented. Section 2 summarises the observations and
Sect. 3 details the data reduction and presents the results. In Sect. 4
we discuss the data and the relevance of these observations for abundance 
determination, as well as the prospects for future such observations with 
8-10m class telescopes.

\section{Observations} 
Long-slit spectra centred on three of the brightest PN in the catalogue 
of Hui et al. (\cite{huib}) were observed with EFOSC1 (Buzzoni et 
al. \cite{buzz}) 
on the ESO 3.6m telescope. Table 1 lists the three PN with the ID numbers,
J2000 coordinates and 5007\AA\ magnitudes (from Hui et al. \cite{huib})
and Fig. 1 shows the positions of the 
PN\#5601 is the brightest PN observed in NGC~5128 (by 0.1mag.).
The targets were chosen to cover a range in galactocentric radius and
where possible the slit orientation was chosen so that at least one 
other PN from the 
Hui et al. (\cite{huib}) catalogue would lie on the slit. For the 
slit length centred on PN\#5601 two other PN were included in the slit
(listed in Table 1);
for PN\#4001 one of the desired PN (4013) was however missed 
by the slit. No target very close to the high surface brightness central 
region of the galaxy was chosen in order to minimize the contribution of 
galaxy continuum to the PN spectra. In total, spectra of five 
PN in Cen-A were detected and 
are all listed in Table 1 together with their radial distance from
the galaxy nucleus. 
Figure 1 shows the positions of these five PN against
the POSS image, with the 1.425 radio contours overlayed
in black (from Condon et al. \cite{cond}).

\begin{figure*}
\resizebox{\hsize}{!}{\includegraphics{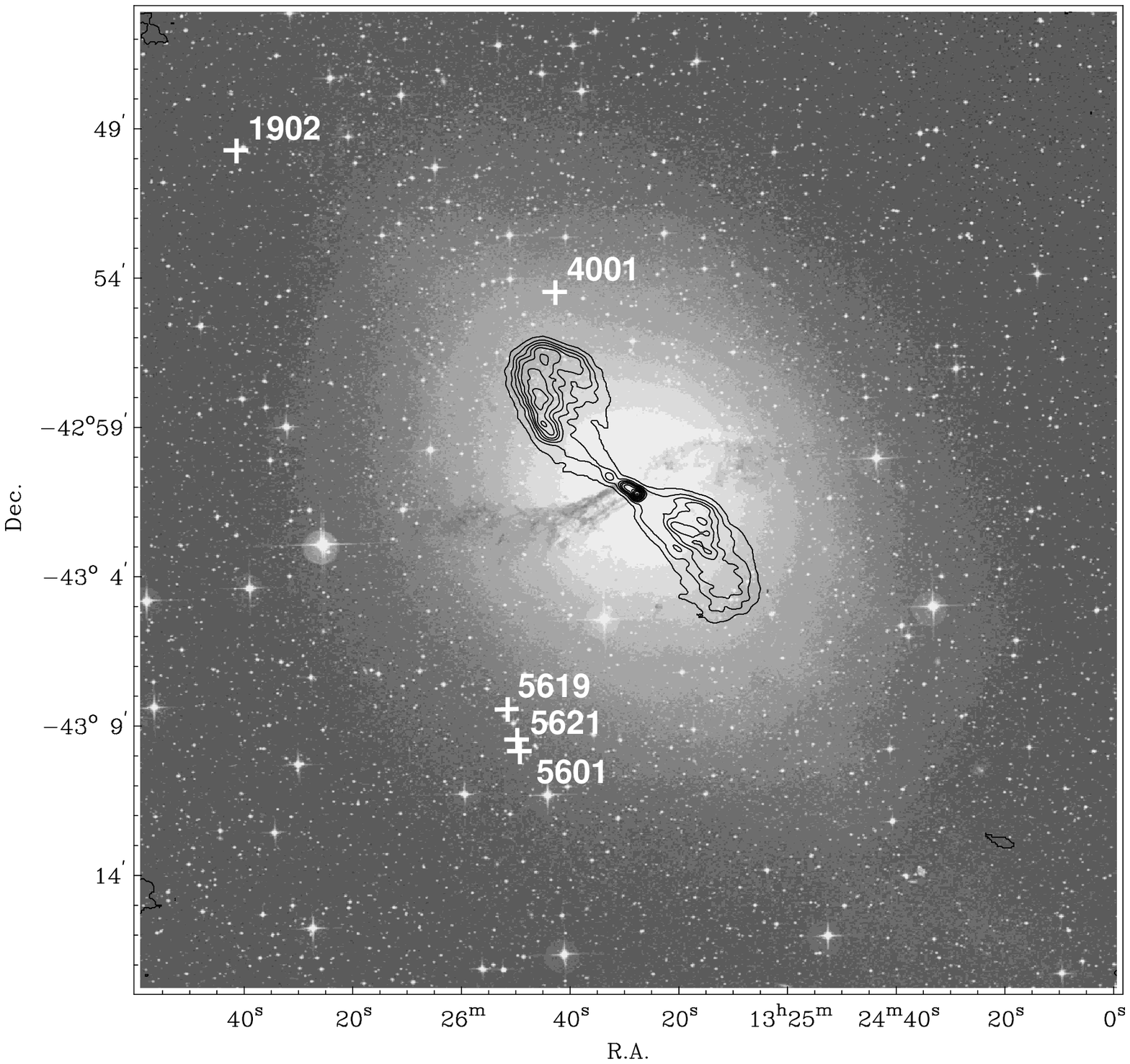}}
\caption{The POSS image of NGC~5128 is shown with the positions of the
five observed planetary nebulae indicated by white 
crosses; the target designations are from Hui et al. (\cite{huib}). 
The contours are from the 1.425 GHz radio continuum map of Condon et al. 
(\cite{cond}). J2000 coordinates are shown.}
\end{figure*}

For each target an [O~III] narrow 
band filter image (ESO\#686, $\lambda_{CEN}$ 5013\AA, $\Delta \lambda$ 
56\AA) and an off-emission filter (ESO\#714, $\lambda_{CEN}$ 5483\AA, 
$\Delta \lambda$ 182\AA, Str\"{o}mgren $y$) image were obtained with 
EFOSC1 in order to acquire the PN on the slit. Exposure times were
usually 5min for the [O~III] image and 2min for the continuum band. 
Visual `blinking' of the two images confirmed the PN and the object was
then centred in the slit. At least one more [O~III] image was obtained
during each sequence of exposures on the same target to ensure that the 
source was well-centred in the slit. 

\begin{table*}
\caption[]{Planetary Nebulae Observed in NGC~5128}
\begin{flushleft}
\begin{tabular}{llcrl}
Central & Target & RA ~(2000)~ Dec & Radial$^{\dag}$ & m$_{5007}$ \\
PN ID   & Name$^{\ast}$   & $h$ ~~$m$ ~~$s$ ~~~~~$^\circ$ ~~$'$ ~~$''$ & 
dist ($''$) & (mag.)$^{\ddag}$ \\
\hline
5601 & 5601 & 13 25 53.52 $-$43 08 54.7 & 547~~ & 23.51 \\
     & 5621 & 13 25 53.81 $-$43 08 37.5 & 535~~ & 25.64 \\
     & 5619 & 13 25 55.31 $-$43 07 12.4 & 474~~ & 25.70 \\
     &      &                         &       \\
1902 & 1902 & 13 26 40.78 $-$42 49 33.6 & 1061~~ & 24.01 \\
4001 & 4001 & 13 25 41.12 $-$42 54 35.0 &  418~~ & 23.89 \\
\end{tabular}
\end{flushleft}
$^{\ast}$ PN designation from Hui et al. (\cite{huib})\\
$^{\dag}$ Measured from the J2000 position of the nuclear radio source
(13$^{h}$ 25$^{m}$ 27.7$^{s}$ $-$43$^\circ$ 01$'$ 06$''$,
Wade et al. \cite{wad}). For reference 1$''$=17pc \\
$^{\ddag}$ $ m_{5007} = -2.5log F_{5007} - 13.74 $,
where $F_{5007}$ is the [O~III]5007\AA\ line flux in ergs cm$^{-2}$ s$^{-1}$
(Ciardullo et al. \cite{cia89}).
\end{table*}

The slit position centred on PN\#1902 also included an emission
line filament (approximate J2000 position 13$^{h}$ 26$^{m}$ 49$^{s}$ 
$-$42$^\circ$ 49$'$ 25$''$). This is part of the system of filaments 
associated with the jet in NGC~5128 some of which have been 
spectroscopically studied by Morganti et al. (\cite{morg}).

  The detector of EFOSC1 was a Tek 512$\times$512 thinned CCD (ESO\#26, 
TK512CB) with 27$\mu$m pixels, which project to 0.61$''$. Given the 
seeing encountered of 1-1.5$''$, slit widths of 1.5$''$ were employed 
to ensure a good balance between receiving the majority of the flux 
from the point-like PN without a severe penalty of sky and galaxy 
background light, and of ensuring
optimal sampling at the detector. The actual slit width employed
is listed in Table 2. All the spectra were obtained with the B300 grism,
which covers the wavelength range 3640 to 6860\AA\ at a dispersion of 
6.3A/pixel; the resulting resolution of the spectra was about 14\AA.
Bias frames, dome flat fields and spectroscopic sky flats were obtained
to correct the CCD pixel response; neon and argon lamps spectra for 
correction of distortions and wavelength calibration; and broad slit
(5$''$) spectra of spectrophotometric standard stars EG~54 (Oke \cite{oke}),
EG~274 (Hamuy et al. \cite{ham}) and LTT~3864 (Hamuy et al. \cite{ham})
for flux calibration.

\begin{table}
\caption[]{Log of EFOSC1 observations}
\begin{flushleft}
\begin{tabular}{cccrc}
Target & Slit width & Date &  Exp. & ZD's \\
        & ($''$)    &      &  (s)~ & ($^\circ$) \\
\hline
 5601  & 1.5  & 1995 Apr 03 & 2400 & 18.3, 14.8, \\
+ 5619 &      &             &      & 15.3, 19.7, \\
+ 5621 &      &             &      & 26.2, 36.0, \\
       &      &             &      & 43.4 \\
       & 5.0  &             &  600 & 30.9 \\
       &      &             &      &  \\
 1902  & 1.5  & 1995 Apr 04 & 2400 & 44.3, 36.8, \\ 
       &      &             &      & 29.5 \\
       &      &             &      &  \\
 5601  & 1.5  & 1995 Apr 04 & 2400 & 19.7, 15.3, \\
+ 5619 &      &             &      & 14.8, 18.5, \\
+ 5621 &      &             &      & 25.9, 34.1, \\
       &      &             &      &  41.4 \\
       &      &             &      &  \\
 1902  & 1.5  & 1995 Apr 05 & 2400 & 43.0, 35.1, \\ 
       &      &             &      & 27.8 \\
       &      &             &      &  \\ 
 4001  & 1.5  & 1995 Apr 05 & 2400 & 21.4, 17.7, \\
       &      &             &      & 14.4, 16.0, \\
       &      &             &      & 21.3, 28.1, \\
       &      &             &      & 35.9, 42.1 \\
\end{tabular}
\end{flushleft}
Resulting total exposure times per target field are: \\
PN\#5601+5619+5621 - 9.50 hrs \\
PN\#1902 - 4.00 hrs \\
PN\#4001 - 5.33 hrs 
\end{table}

The slit orientation was kept fixed for each set of observations of a given 
target. This was necessary to avoid the delays resulting from many
replacements of the slit on the target which would have resulted from 
the conventional tracking of the parallactic angle by slit rotations.
In addition, keeping a fixed slit orientation facilitated the subtraction
of the underlying (galaxy) continuum and the detection of several PN along the
slit lengths. However imposing a fixed orientation leads to differential
loss of light with wavelength, largest at the higher zenith distances, as 
the parallalactic angle differs from the slit position angle. The requirement
of long integration times and only three allocated nights forced us to observe
the targets for some (small) fraction of the time at zenith distances exceeding 
40$^\circ$, where the differential 
atmospheric refraction between 3700 and 6700\AA\ exceeds the slit width
(e.g. Fillipenko \cite{fill}). 
In order to attempt to control the amount of wavelength-dependent slit loss through
differential refraction, we adopted an observing strategy of always
including a bright stellar source on the slit. The spectrum of this source
could then be used to monitor, and correct, the differential 
refraction losses.
Comparison of the extracted spectra of the star, corrected for atmospheric
extinction, at high and low zenith distance should allow this correction 
to be applied
to the spectra of the PN. One exposure of PN\#5601 with the star on the slit 
was also made with a broad slit (5$''$) in order to test the validity of this 
technique with a spectrum essentially free from any differential slit losses.
  
\section{Reductions and Results}
\subsection{Reductions}
All the spectra were reduced in the usual way, using the
spectroscopic packages in IRAF. A super-bias frame was formed by averaging
many individual bias frames and this
bias image was subtracted from all frames. A mean flat field image was formed
from many exposures to a tungsten
lamp shining off a reflector in the dome and was used to rectify 
the pixel-to-pixel variations. The sky flat was employed to map the 
response of the system in the cross dispersion direction, thus correcting
any vignetting in the optical system or variation in slit transmission. 
The exposures of the Neon and
Argon lamps were employed to fit the known wavelengths of the
comparison lines and the spectra were rebinned into
channels of constant wavelength width by fitting third
order polynomials. Figure 2 shows the long slit spectrum centred on PN\#5601
formed by averaging (with cosmic ray rejection) the first four exposures 
on April 03 (Table 2). The other two PN (5619 and 5621) are clearly 
visible from their [O~III] line emission. In addition there are a number 
of continuum sources detected, some of which are probably stellar 
clusters in NGC~5128. There is a faint red star displaced less than 
one seeing disc from the PN\#5601 and visible in Fig. 2. The bright 
continuum source closest to PN\#5619 was the one used in attempting to 
correct for differential refraction slit losses.

\begin{figure*}
\resizebox{\hsize}{!}{\includegraphics{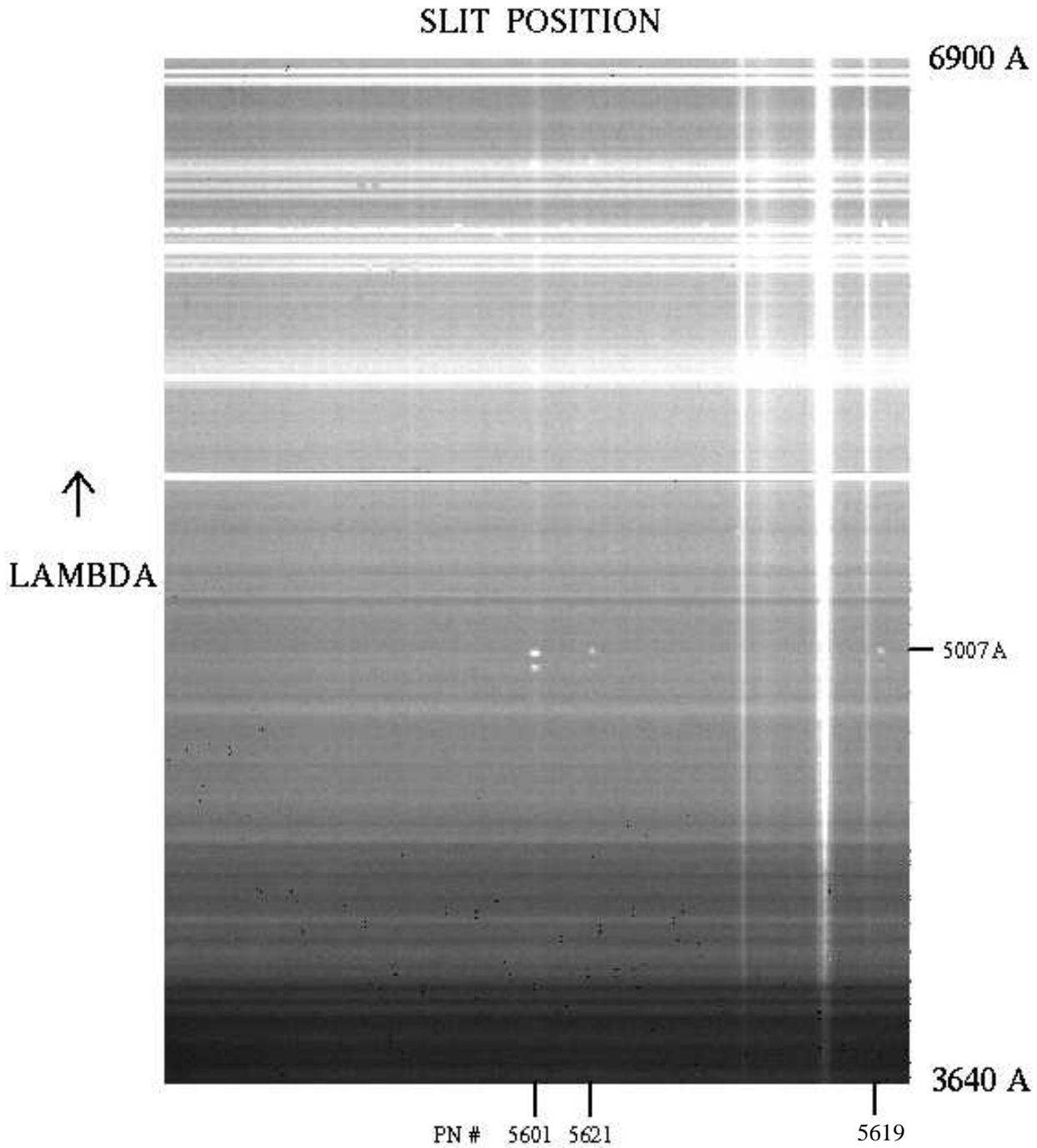}}
\caption{The 2-D longslit spectrum centred on PN\#5601 (see
Table 1) formed by averaging the first four exposures on April 03 
(see Table 2) is shown. Wavelength increases from bottom to top and 
the left edge corresponds to south.  The position of the 
[O~III]5007\AA\ line is indicated as are the positions of the three 
detected PN (see Table 1).}
\end{figure*}

Spectra of the PN, and the reference star, were extracted from each
image using 
optimal weighting after subtracting the mean sky from the vicinity of the 
objects. The atmospheric
extinction was corrected for each exposure and absolute flux calibration 
applied from the observation of the spectrophotometric standard.
The background contained a substantial contribution of
galaxy light, especially for PN\#4001 so it was necessary to
restrict the background region to be close to the PN (with
typically at least three times as many pixels in the sky as in the 
extracted PN). After extraction and removal of cosmic rays,
the individual spectra were flux calibrated. 

  The spectra of the reference continuum source from each exposure
were compared. The spectra showed a very large difference.
Spectra taken before transit of the star indicated an upward  
correction to the blue fluxes and a downward correction to red fluxes
relative to the spectrum taken at the lowest airmass.
For the spectra taken after transit of the star a downward  
correction to the blue fluxes and an upward correction to red fluxes,
relative to the spectrum taken at the lowest airmass, was found.
However applying such corrections to the extracted PN spectra gave 
inconsistent line fluxes in the sense that the correction 
factors were too steep with wavelength and resulted in discordant
spectra from before and after meridian passage. The explanation for the
derivation of such unrealistically large corrections is unclear.
The continuum source could be extended (e.g. be a cluster in NGC~5128
itself) and have different colours in different regions; however this
seems unlikely since exactly the same behaviour was exhibited by
the reference spectra for the other targets. The most probable
explanation is that the slit rotates slightly during the course of the
observations so that the flux received in the slit tracks across the
image in opposite directions on either side of the meridian, 
exaggerating the effects of differential atmospheric extinction.
It was found that the line ratios of the extracted (and extinction
corrected) PN spectra did not vary systematically with airmass
beyond the errors of measurement.
In addition the line ratios in the extracted spectra did not differ 
from those of the broad slit exposure of PN\#5601 by more than the 
errors, although flux determination of the H$\alpha$ line was 
hampered by the broadened sky lines. Since emission lines were only 
detected over the wavelength range 4500 to 6700\AA, and the 
differential atmospheric refraction is 0.70$''$ over this range
for an airmass of 1.4, then in 1.5$''$ seeing with a 1.5$''$ slit
the differential flux loss was at maximum 30\% (see Fig. 1 
of Jacoby \& Kaler \cite{jaka}). Thus overall only small losses 
in spectrophotometric 
integrity of the combined spectra should result. The fluxed spectra 
for each PN were averaged (using weights based on exposure time)
on a case by case basis excluding the
last exposure at highest airmass to form the final PN spectra. 
The extracted spectrum of the jet filament was treated similarly.
 
\subsection{Results}
Figure 3 shows the mean spectra of the five PN observed in Cen-A.
The red star continuum under PN\#5601 could not be effectively subtracted.
The emission lines were interactively fitted by Gaussians and the
flux in the lines are listed in Tables 3 and 4 for each PN and the
filament. In Table 3 the line flux data for the three brighter PN 
are collected. The errors in Table 3 take into account the continuum 
under the line and the photon noise in the sky-subtracted spectra;
the errors on the H$\beta$ flux have been propagated to the
other line flux errors. The measured signal-to-noise on the 
[O~III]5007\AA\ line flux for the brightest PN (5601) is 55.
The reddening correction was calculated by comparing the observed 
H$\alpha$/H$\beta$ ratio to the Case B value using the Seaton 
(\cite{seat}) Galactic reddening law and is listed in Table 3. 
The dereddened line fluxes (the error on the extinction 
was not propagated to the dereddened line errors) together 
with the observed H$\beta$ flux are listed in Table 3. In Table 4 
the fluxes are presented for the two fainter PN and the filament 
near PN\#1902. The errors are substantially larger than for the 
data presented in Table 3, since the PN are fainter; for 
example for PN\#5619, the signal-to-noise on the measurement of the 
[O~III]5007\AA\ line is 9.

\begin{figure*}
\resizebox{\hsize}{!}{\includegraphics{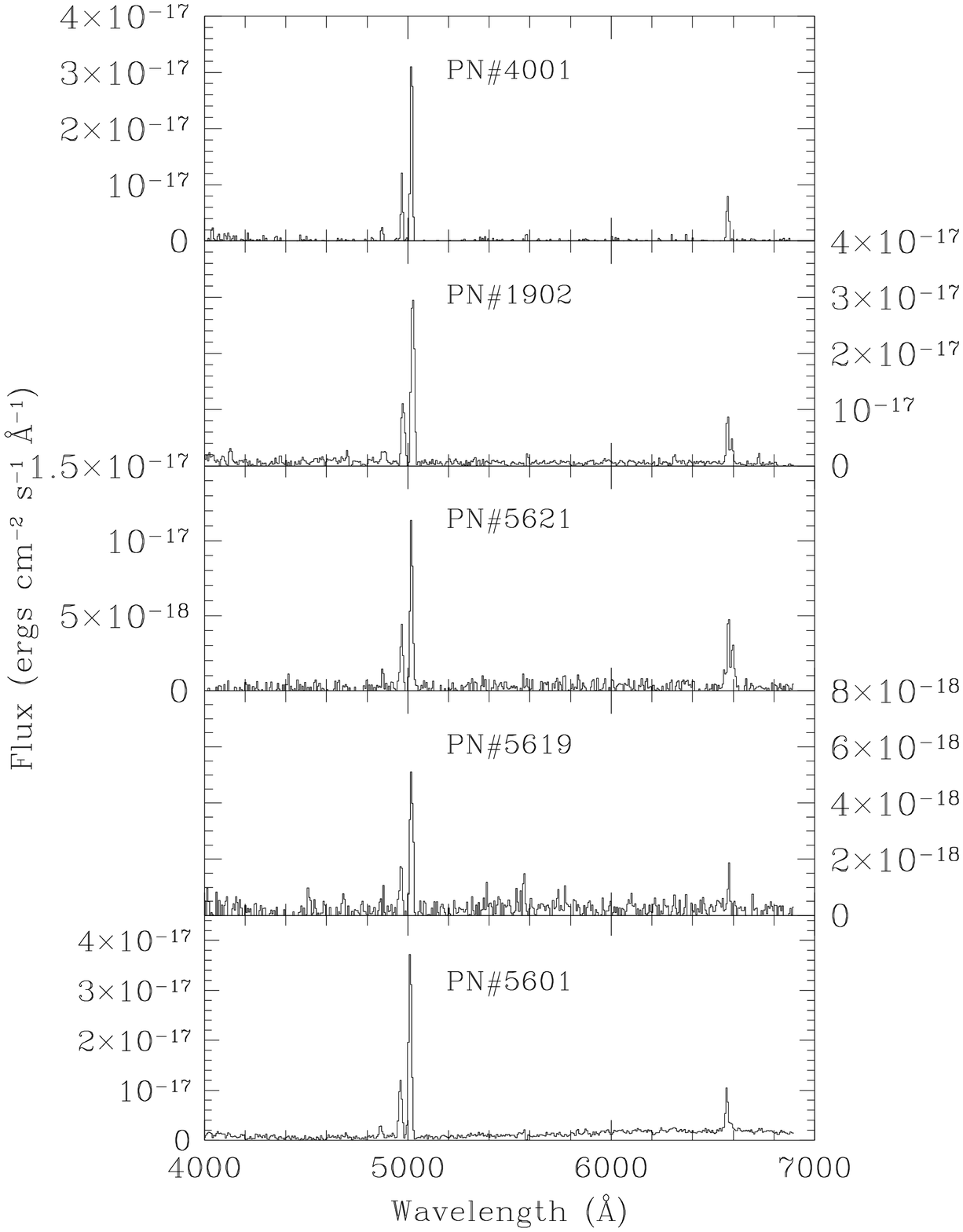}}
\caption{The spectrum of the five PN detected in
NGC~5128 is shown. The designation of each object is indicated
(from Hui et al. \cite{huib}).}
\end{figure*}

\begin{table*}
\caption[]{Emission line fluxes of three brightest PN observed in NGC~5128} 
\begin{flushleft}
\begin{tabular}{cl|rrrr|rrrr|rrrr}
  PN\# &    & \multicolumn{4}{c}{1902} & \multicolumn{4}{c}{4001} & \multicolumn{4}{c}{5601} \\
Ident. & $\lambda$ & I$_O^\ast$ & $\pm$ & I$_D^\dag$ & $\pm$ & I$_O$ & $\pm$ & I$_D$ & $\pm$ & I$_O$ & $\pm$ & I$_D$ & $\pm$ \\
       & (\AA)     &       &       &       &       &       &       &       &       &       &       &       & \\
\hline
He~II & 4686 & 64 & 23 & 63 & 23 & 12 & 3 & 12 & 3 & 20 & 16 & 20 & 17 \\
H$\beta$ & 4861 & 100 & 0 & 100 & 0 & 100 & 0 & 100 & 0 & 100 & 0 & 100 & 0 \\
{[O~III]} & 4959 & 498 & 93 & 502 & 93 & 330 & 76 & 331 & 76 & 414 & 51 & 416 & 51 \\
{[O~III]} & 5007 & 1185 & 216 & 1197 & 219 & 960 & 218 & 965 & 219 & 1268 & 153 & 1268 & 153 \\
He~I & 5876 & 31 & 7 & 33 & 7 & & & & & 30 & 16 & 31 & 17 \\
H$\alpha$ & 6563 & 261 & 49 & 286 & 54 & 274 & 57 & 286 & 62 & 271 & 39 & 286 & 41 \\
{[N~II]} & 6583 & 135 & 28 & 147 & 30 & 18 & 5 & 17 & 5 & 56 & 11 & 59 & 12 \\
{[S~II]} & 6716+31 & 58 & 27 & 63 & 30 & & & & & 24 & 9 & 25 & 10 \\
\hline
c        &         & \multicolumn{4}{c}{-0.12$\pm$0.23} & \multicolumn{4}{c}{-0.06$\pm$0.28} &
\multicolumn{4}{c}{-0.07$\pm$0.19} \\
Log$_{10}$F(H$\beta$)$^\ddag$ & & \multicolumn{4}{c}{-16.3} & \multicolumn{4}{c}{-16.4} & \multicolumn{4}{c}{-16.2} \\ 
\end{tabular}
\end{flushleft}
\noindent
$^\ast$ Observed flux normalised to I(H$\beta$)=100 \\
$^\dagger$ Dereddened flux normalised to I(H$\beta$)=100 \\  
$^\ddag$ Absolute observed H$\beta$ flux
\end{table*}

\begin{table*}
\caption[]{Emission line fluxes of fainter PN and filament observed in NGC~5128} 
\begin{flushleft}
\begin{tabular}{cl|rrrr|rrrr|rrrr}
  PN\# &    & \multicolumn{4}{c}{5619} & \multicolumn{4}{c}{5621} & \multicolumn{4}{c}{Filament} \\
Ident. & $\lambda$ & I$_O^\ast$ & $\pm$ & I$_D^\dag$ & $\pm$ & I$_O$ & $\pm$ & I$_D$ & $\pm$ & I$_O$ & $\pm$ & I$_D$ & $\pm$ \\
       & (\AA)     &       &       &       &       &       &       &       &       &       &       &       & \\
\hline
He~II & 4686       &      &     &       &     &  46 &  38 &  48 &  38 &      &     &      &     \\
H$\beta$ & 4861    & 100  &   0 &  100  &   0 & 100 &   0 & 100 &   0 &  100 &   0 &  100 &   0 \\
{[O~III]} & 4959   & 482  & 190 &  484  & 190 & 365 & 153 & 358 & 150 &  452 & 136 &  441 & 133 \\
{[O~III]} & 5007   & 1484 & 480 & 1489  & 482 & 904 & 360 & 879 & 355 & 1074 & 313 & 1036 & 302 \\
H$\alpha$ & 6563   & 278  &  96 &  286  &  99 & 365 & 127 & 286 & 100 &  382 & 113 &  286 &  85 \\
{[N~II]} & 6583    &      &     &       &     & 252 & 100 & 197 &  80 &  251 &  73 &  186 &  54 \\
{[S~II]} & 6716+31 &      &     &       &     &  67 &  30 &  51 &  23 &      &     &      &     \\
\hline
c        &         & \multicolumn{4}{c}{-0.04} & \multicolumn{4}{c}{0.33} &
\multicolumn{4}{c}{0.40} \\
Log$_{10}$F(H$\beta$)$^\ddag$ & & \multicolumn{4}{c}{-17.1} &
\multicolumn{4}{c}{-16.7} & \multicolumn{4}{c}{-16.6} \\ 
\end{tabular}
\end{flushleft}
\noindent
$^\ast$ Observed flux normalised to I(H$\beta$)=100 \\
$^\dagger$ Dereddened flux normalised to I(H$\beta$)=100  \\
$^\ddag$ Absolute observed H$\beta$ flux
\end{table*}

\section{Discussion}
\subsection{Planetary Nebula spectra}
The five PN observed in NGC~5128 show spectra entirely typical of PN;
the spectra are not obviously distinguishable from those of Galactic PN.
Although the signal-to-noise is not high, the range of line fluxes,
absolute H$\beta$ fluxes and [O~III]5007\AA/H$\beta$ ratios is similar
to that for high excitation Galactic PN. There were no low excitation
PN spectra among the five, but this is probably not surprising given that 
the source detection was performed in [O~III] and the emphasis here was
on the brightest objects.  The range in [O~III] 
brightness covered is 7.5 (from the photometry of  Hui et al. \cite{huib}) 
and 8 in H$\beta$ flux from the long slit observations (Tables 3 and 4).  
From the standpoint of the evolution of low mass stars it is not
surprising that the spectra are typical, but on the other hand
these are among the brightest PN in a whole galaxy. Doubt had been expressed
that the PN at the peak of the luminosity function may not have been
typical of the general PN population and that distance estimates which 
relied on the peak of the luminosity function could suffer from 
systematic bias (Bottinelli et al. \cite{bott}; Tammann, \cite{tam}). 
These effects have been carefully refuted (Feldmeier et al. \cite{feld},
McMillan et al. \cite{mcmi} and Jacoby \cite{jac96}) and together
with the spectra shown in Fig. 3 and tabulated in Tables 3 and 
4 amply demonstrate that the brightest PN in a galaxy {\bf are} typical. 
That they are the brightest is simply due to the fact that they are
observed whilst at their peak luminosity. The 5007\AA\ luminosity is 
generally higher for higher mass progenitor stars and also peaks in the 
later stages of evolution of lower mass stars (Sch\"{o}nberner \& Tylenda 
\cite{scho}). However for PN with high core masses, above $\sim$0.65M$_\odot$,
the high nitrogen abundance can decrease the efficacy of cooling by oxygen
emission, reducing the 5007\AA\ flux by $\sim$ 0.5mag (Kaler \&
Jacoby \cite{kaja}). 

  The extinction correction for the three brightest objects indicates 
a slightly negative value. The Galactic component of reddening to 
NGC~5128 given by Burstein \& Heiles \cite{buhe} is E$_{B-V}$ = 0.123 
(c=0.18). This value is similar to that (E$_{B-V}$ = 0.10) adopted by 
van den Bergh (\cite{berg}) and that derived most recently from 
DIRBE dust maps
by Schlegel et al. \cite{schl} (E$_{B-V}$ = 0.115); however 
Jablonka et al. (\cite{jabl})
measured values of E$_{B-V}$ as low as 0.03 from spectrophotometry
of globular clusters in NGC~5128. The extinction values for 
the PN are consistent with these values within the errors, except 
perhaps for PN\#5601. However it is puzzling 
that the values are systematically low; any local extinction in
NGC~5128, or dust within the nebulae themselves, would increase the
value above the baseline for the Galactic line of sight extinction.
Slit lossses through atmospheric refraction should not alone
account for bias. Errors may have arisen in subtraction of the
underlying stellar continuum whereby emission line flux is lost
to stellar absorption lines, although the effect would be to produce 
an increased extinction on account of the generally higher H$\beta$ 
absorption equivalent width compared with H$\alpha$. However one of the
PN (5621) does show an extinction above the Galactic value, although
with a substantial error, as does the filament, which is however an
extended object. The extinction to
the filament is in the range of values for the extinction determined
by Morganti et al. (\cite{morg}); the filament is in the same
vicinity as their Field 2 (see their Fig. 2) although about ten 
times lower 
surface brightness (compare their Table 4 for spectra). The 
[O~III]5007\AA/H$\beta$ ratio measured here is also similar to
the values measured by Morganti et al. (\cite{morg}).

The most probable explanation of the depressed reddening values
is that the central wavelength of the guiding camera 
lies at one end of the range H$\beta$ to H$\alpha$; by tracking on
the image in the vicinity of the wavelength around H$\beta$, flux is
systematically lost from the slit at H$\alpha$ for airmasses much
greater than 1.0. Subsequent to this conclusion we were informed
that the guiding camera of the ESO 3.6m at the time of the
observations was sensitive over the wavelength range 3700 to 5000\AA. 
The dereddened fluxes were formed employing the observed reddening, 
even if negative; this serves to compensate for the losses of the
red part of the spectra. No specific correction for foreground 
(Galactic) reddening was employed. 
It is apparent that the three brightest PN show no evidence for
intrinsic reddening (within the substantial measurement errors), which 
is probably not surprising given that they are the the brightest PN
observable in the galaxy. Any extinction would move them to 
lower observed fluxes; PN\#5621 is for example as intrinsically
bright as PN\#4001. The effect of local galactic extinction and
dust intrinsic to the PN must play a role in shaping the PN
luminosity function (Jacoby \cite{jac89}).
With spectroscopy of the brightest PN, the effect of dust on the 
luminosity function, and hence on the distance estimate through 
fitting of this function (Ciardullo et al. \cite{cia89}; see also
Mendez et al. \cite{men}), can be directly quantified.
However if dust is associated with PN dependent on their
luminosity it would be expected to have a strong effect on the PN 
luminosity function. Jacoby \& Ciardullo (\cite{jaccia}) in their
study of PN in M~31 have found a weak correlation between extinction
and 5007\AA\ luminosity, which also exists for PN in the LMC. The
surprising net effect on the PN luminosity function is that the apparent 
peak brightness is nearly independent of absolute peak brightness.

On the basis of the [O~III]5007\AA/H$\beta$ and He~II~4686\AA/H$\beta$
ratios the excitation class can be defined (Dopita \& Meatheringham
\cite{dom90}). A least squares fit of the effective temperature
from photoionization models (Dopita \& Meatheringham \cite{dom91b}) 
against excitation class for a uniform set of observations and 
models of Magellanic Cloud PN allows estimation of effective temperature.
Use of this data set could  be criticized since the LMC and SMC have
low metallicities compared to Galactic or more metal rich
galaxies, but the modelling of
AGB evolution shows no strong dependence of stellar temperature on
metallicity (Dopita et al. \cite{djv}). 
Table 5 lists the excitation class and indicative temperature of 
the PN in Cen-A. PN\#5619 could not have a reliably assigned 
stellar temperature since its excitation class is high (based on
its high [O~III]/H$\beta$ ratio) yet it was too faint to detect 
He~II4686\AA\ (excitation class above 5.0 requires the He~II/H$\beta$ 
ratio). These temperatures should be seen as upper limits, since 
if the nebulae are optically thin the high ionization emission
is enhanced; given the large line ratio errors the likely errors 
are at least $\pm$10000K. 

\begin{table*}
\caption[]{Parameters of the NGC~5128 Planetary Nebulae}
\begin{flushleft}
\begin{tabular}{lrrrrr}
Parameter & PN\#1902 & PN\#4001 & PN\#5601 & PN\# 5619 & PN\# 5621 \\
\hline
Excit. Class  & 7.8 & 5.0 & 5.4 & 6.7 & 7.0 \\
T$_{\ast}$(K) & 180000 & 100000 & 110000 & $\sim$140000 & 155000 \\
Log L$_{\ast}$ & 4.0 & 3.9 & 4.2 & - & 3.7 \\
M$_{\ast}$ (M$_\odot$) & 0.68 & 0.64 & 0.83 & - & 0.62 \\
T$_e$(K) & 13000 & 14000 & 13000 & 14000 & 12500 \\
Z(Z$_\odot$) & -0.3 & -0.6 & -0.4 & -0.6 & -0.4 \\
O$^{++}$/H $\times$10$^{5}$ & 26 & 13 & 21 & 14 & 25 \\
12$+$Log$_{10}$(O/H) & $<$8.5 & 8.2 & 8.5 & $>$8.4 & $<$8.3 \\
$[O/H]$ & $>$-0.4 & -0.7 & -0.4 & $<$-0.5 & $>$-0.6 \\
N/O & 0.4 & 0.3 & 0.4 & - & 0.5 \\   
\hline
\end{tabular}
\end{flushleft}
\end{table*}

\subsection{Abundances of the Planetary Nebulae}
Even for the three brightest PN observed in NGC~5128, the weak
diagnostic forbidden lines were not detected; thus it is not
possible to measure accurate electron temperatures (from the 
[O~III]5007/4363\AA\ ratio) or densities (for example from the 
[S~II]6716/6731\AA\ ratio). Nevertheless an attempt was made to
estimate the oxygen abundance in order to compare it with other
abundance diagnostics (e.g. from stellar absorption lines).
Dopita et al. (\cite{djv}) presented a diagnostic diagram of
PN metallicity {\em v.} effective temperature, in which the
electron temperture can be determined from the [O~III]/H$\beta$
ratio. This plot was derived from photoionization models
of a grid of optically thick PN; the effective temperature
being determined from the fit to the Magellanic Cloud data
(Dopita \& Meatheringham \cite{dom91b}).
In Table 4 the estimated values of the electron temperature
are listed; for PN\#1902 and 5621 the range of effective
temperatures are out of range of the diagnostic plot, but
were extrapolated. Row 6 of Table 4 lists the 
derived metallicities of the PN (all element abundances 
scaled except Helium), based on the Dopita et al. (\cite{djv}) 
calibration. From the electron temperature estimates (Table 4 
row 5), the empirical O$^{++}$ abundances were determined
from the [O~III]/H$\beta$ ratios and are listed in row 7.
A correction for the presence of O$^{3+}$ was made using the
ionization correction factor derived from the He/He$^{++}$ 
ratio (Kingsburgh \& Barlow \cite{kiba}); the He/H ratio 
was assumed fixed at 0.15. The He~I 5876\AA\
line indicates He$^{+}$/H$^{+}$ ratios as high as 0.2 but 
this line suffers from proximity to the strong Na~I telluric
lines, and the difficulty in good sky subtraction leads to
a large error on the line measurement. Assuming that the N/O ratio 
for the PN is the same as the mean value for the Galactic PN
(0.28; Kingsburgh \& Barlow \cite{kiba}), the fraction of
O$^{+}$/H$^{+}$ was estimated and thus the total oxygen 
abundance.  The uncertainties introduced in 
correcting for the presence of O$^{3+}$ and O$^{+}$ are 
between  20 and 40\%. Where the [N~II]6583\AA\ line was 
strong it was assumed that the N/O ratio was higher than the 
Galactic mean value and the O$^{+}$
contribution was included as an upper limit. Row 8 lists the 
derived logarithmic O/H abundances and row 9 the oxygen 
abundances compared to solar.

  In addition to these empirical estimations of the nebular parameters,
the photoionization modelling package CLOUDY (Ferland \cite{ferl})
was used to model the spectra matching the [O~III] luminosity and
relative line strengths; the carbon abundance was assumed as 
12$+$Log$_{10}$(C/H)=8.7. The derived parameters were generally in
good agreement with those empirically derived; the stellar temperatures
were about 10000K lower however. 
The fair agreement between the
empirical O/H abundance estimates and those from photoionization 
models in Table 5 provides assurance that the use of empirical relations 
calibrated from lower metallicity Magallanic Cloud planetary nebulae
does not seriously affect (within $\sim$0.15dex) the resulting
abundance estimates. 
In rows 3 and 4 of Table 3, the derived 
stellar luminosity and core mass (from Sch\"{o}nberner \cite{sch81},
\cite{sch83} tracks) are also listed. From the models, the diagnostically 
useful N/O ratio was calculated and is listed in row 10. Due to the limited 
spectroscopic constraints on the models, the nitrogen abundances are 
uncertain; the N/O ratios could be a factor of 2 -- 3 higher, but are not 
likely to be much lower if the PN are optically thick (which appears 
probable both from the photoionization models and their high luminosity).
Whilst the values
of N/O are moderately high (the mean value for non-Type I Galactic PN is
0.28 - Kingsburgh \& Barlow \cite{kiba}), only PN\#5621 satisfies the
criterion of N/O $>$ 0.5 for classification as a Type I PN,
considered to arise from higher mass progenitor stars (Peimbert \&
Torres Peimbert \cite{pepe}). This object in addition displays
He~II emission strong relative to H$\beta$, 
so may be a bona fide 
Type I nebula (He/H$\geq$0.125). A direct measurement of the
N/O ratio (such as from [N~II]6583/[O~II]3727\AA\ line ratio) would be 
required to confirm this classification. However the two brightest
PN (5601 is the brightest PN in the galaxy detected by Hui et al.
\cite{huia}) are not obviously Type I PN; this is consistent with 
observations of PN in the Magellanic Clouds (Dopita \& Meatheringham,
\cite{dom91b}) that Type I PN are not the most luminous in a 
population on account of their fast evolution to high effective 
temperatures and hence lower luminosities. Type I PN could be more 
optically thin than their lower mass counterparts as suggested by 
Mendez et al. (\cite{men}), thereby leading to lower observed [O~III] 
luminosities than expected from the luminosities of the central stars.
Type I PN may also be more copious producers of dust, hence further
lowering their observed luminosities. 

 The quality of the oxygen abundance determinations is not high
enough to investigate any evidence of a metallicity gradient;
the O/H abundance {\em v.} projected radius shows no trend, perhaps 
even a suggestion of increasing with increased galactocentric radius.
The sample is too small and the quality of the O/H determinations
too low to draw any conclusions. The mean [O/H] abundance appears
to be -0.5 for the PN in Cen-A; this can be compared with the
mean value for 42 non-Type I Galactic PN of -0.24 (Kingsburgh \& Barlow 
\cite{kiba}). It is surprising that the oxygen abundance 
is lower than characteristic for Galactic PN, given that the 
metallicity is expected to be higher in this high luminosity elliptical
galaxy. The range of galactocentric radii
probed by the five PN is however 7.1 to 18.0 Kpc (Table 1), which, 
by analogy with the Milky Way would show a lower metallicity
than the core (by [O/H] $\sim$ -0.3 at 7Kpc e.g. Shaver et al.
\cite{shav}).  

There does not appear to be a direct metallicity determination for the
large scale stellar content of this galaxy.
However the metallicity of NGC~5128 can be estimated using the 
tight relations between velocity dispersion, luminosity, and 
Mg$_2$ index for giant ellipticals (Faber \& Jackson \cite{faja};
Terlevich et al. \cite{terl}). The Mg$_2$ index can then be 
converted to metallicity using stellar population models. 
NGC~5128 has a central velocity dispersion between 150 and 200 km s$^{-1}$
(Wilkinson et al. \cite{wilk}). This might however be a lower limit, 
since the inner regions are obscured by the large central dust lane. 
Wilkinson et al. (\cite{wilk}) conclude that M$_B \approx -$20.5 mag.;
this then corresponds to an Mg$_2$ index of 0.31. To convert this 
value into a metallicity, a value for the age of NGC~5128 must be 
assumed. For an age of 17 Gyr Vazdekis et al. (\cite{vaz}) give 
a metallicity slightly higher than solar, or [Fe/H]=0.4 for an age of 
6 Gyr. These numbers again depend slightly on the IMF chosen, but it 
is fairly certain that the stellar indicators show an abundance 
which is about solar or higher. In addition the globular clusters 
have a higher mean metallicity than for the Milky Way. In NGC~5128 
Harris et al. (\cite{har92}) determined a mean metallicity 
$<[$Fe/H$]>$ of -0.8 for 62 globulars from Washington photometry
whilst the mean for all globulars in the Milky Way is  -1.35. Clearly
a determination of the stellar metallicity variation with radius is
required for NGC~5128 to contrast with the measurements from the
planetary nebulae. It should also be a priority
for future observations to study PN near the galaxy core (but 
avoiding the dust lane) in order to search for high abundance PN.

 A detailed investigation of the discrepancy between the stellar 
and PN abundances is beyond the scope of this paper involving as
it does stellar evolution, chemical enrichment processes, mergers
and elliptical galaxy formation. However a similar discrepancy
between the stellar and PN metallicity
is seen with abundance data from PN in the bulge of M~31, where
a mean [O/H]$\sim$-0.5 is found (Jacoby \& Ciardullo \cite{jaccia}) in strong
contrast to the apparently super-solar stellar abundances.
One obvious reason for such discrepancies could be that the PN and
stellar abundances do not refer to the same stars; for example the
absorption line spectra would be weighted by the most luminous
stars. In the optical the more metal rich stars are generally fainter
than lower metallicity ones so that the stellar indicators 
(e.g. Mg$_{2}$ index) are weighted to older stars, except
in young populations (less than a few Gyr). However
if the chemical enrichment
proceeds monotonically with time, then the younger stars are
more metal rich; although enrichments of 0.5 dex in a few Gyr might
require epochs of star formation rather than steady evolution.
If mergers with lower luminosity (and metallicity) galaxies 
contributed substantially to the stellar population this would 
decrease or even reverse the trend of increasing metallicity
with time. NGC~5128 appears to have suffered a recent merger. 
Unlike most other ellipticals it has a prominent twisted disk of 
gas containing numerous HII regions, and lying approximately along 
the galaxy minor axis. The velocities of the PN (Hui et al. 
\cite{huic}) indicate that the inner disk, containing the HII 
regions, is rapidly rotating or is the remnant of a precessing, 
nearly polar, gas disk in an axisymmetric potential (Sparke 
\cite{spark}). The five PN observed could have originated from 
the smaller, lower metallicity, infalling galaxy; however only 
a fraction of the total number of PN in NGC~5128 could have been 
so produced since the luminosity specific PN density for NGC~5128 
is similar to that for other early-type galaxies (Hui et al. 
\cite{huia}).

A second reason for the low O abundance from the PN in comparison
with the expected high stellar (Fe peak) abundance is suggested
by the known anti-correlation between (Galactic) stellar Fe/H and O/Fe
(King \cite{king}). If the stellar population is super metal
rich when considered from the viewpoint of Fe abundances it
is not from the perspective of O abundances, which the PN
reflect, independent of any stellar O/Fe calibration. 
However such a relation may be rather specific to
the chemical evolution history of the Galaxy which must be
very different from that of NGC~5128. But ellipticals
tend to show Mg/Fe larger than solar (e.g. Worthey et al. \cite{wor})
and the stellar Mg abundance should follow O (Faber et al. 
\cite{fab92}) which brings back the discrepancy between the PN and 
the stellar abundances. The extensive data
on PN in different galaxies collected by Stasinska et al. (\cite{stas})
show that O/H deduced from the PN is dependent on their luminosity
(viz. stellar core mass), being larger for high luminosity PN.
In addition the more luminous PN may be younger and should then 
probe the
interstellar medium at later epochs. Although their sample does
not encompass massive early-type galaxies, it implies that the
low mean O/H of the five PN in NGC~5128 is difficult to understand.
The luminosity specific PN number density is also high for 
NGC~5128, consistent with the bluer colour of NGC~5128 
(Peimbert \cite{peim}; Hui et al. \cite{huia}) suggesting
a lower metallicity than the giant ellipticals in Virgo, 
such as NGC~4472. Population age may also be a consideration for 
PN number density, since it appears from
the relatively few number of PN in Galactic globular clusters
that old stars produce relatively few PN (Jacoby et al. \cite{jac97}).

Contamination of the PN sample by halo objects, which, by 
comparison with the Galactic PN sample (e.g. Howard et al. 
\cite{how}) have notably low metallicity, could be reduced by 
considering the kinematics. From the NGC~5128 PN
kinematic survey (Hui et al. \cite{huic}), PN\#5601 and 
1902 could be halo objects but their O abundances are on the
high side of the mean (Table 5). The effect of high metallicity
on PN formation is also not known - for example an 
enhanced mass loss rate on the AGB could result in dispersal of the
envelope before the central star has had time to heat up enough
to ionize the nebula (the AGB manqu\'{e} channel Greggio \& 
Renzini \cite{gren}; see also Ferguson \& Davidson \cite{feda})
This phenomenon would give rise to fewer PN at higher metallicity
explaining the lower average metallicity of the PN. It is expected
that the brightest PN come from a slightly metal poor
population (Ciardullo \& Jacoby \cite{cija}) as predicted
by the models of Dopita et al (\cite{djv}). In this
context it would be useful to search for Galactic bulge PN 
with super-solar metallicity in order to reach a clearer
understanding of the role of metallicty on PN evolution.
In addition more and better data on the
PN and the stellar populations in metal rich systems, not
confined to the bulges of spirals, are required for an 
understanding of what controls PN evolution in such environments
before detailed evolutionary scenerios in particular galaxies
can be developed.

\subsection{Exploring Abundance Gradients with PN}

In order for the PN to be reliable tracers, their abundances
must reflect those of the gas from which the stars were formed
and not solely be a consequence of nuclear reprocessing. From studies
of Galactic PN, the O, Ne, S and Ar gradient (Maciel \&
K\"{o}ppen \cite{mac}) matches that of the H\,II
regions (Shaver et al. \cite{shav}) as does the He gradient (Peimbert
\& Serrano \cite{pei}). This applies to the
(common) Type~II nebulae, not to the minority Type~I PN,
originating from higher mass progenitors and having enhanced
He, N and Ne. The Type II PN and H\,II regions in the lower
metallicity environment of the Magellanic Clouds also indicate
similar abundances (see Clegg \cite{cleg}). Richer
(\cite{rich}) has arrived at the important conclusion
that the brightest PN in ellipticals have the same status as
abundance indicators as H\,II regions in spirals. Spectra 
comparable to or better than the ones presented here for 
PN\#1902, 4001 and 5601 are required to distinguish the Type II
from Type I nebulae and to determine  improved oxygen abundances. 
For the brighest PN it will be feasible to detect the [O~III]4363\AA\ 
line and thus determine O abundances to $\pm$0.2dex or better;
for lower luminosity PN, or galaxies more distant than NGC~5128,
empirical abundance determination and photoionization modelling, 
as performed here, is required
and the derived abundances are of lower individual weight. However
average oxygen abundances at least comparable in accuracy to
those for the integrated stellar population as a function of
radius, and for individual globular clusters, are achievable.

  A rather large sample of PN is required to distinguish a
trend in the abundance, since a PN at a given effective radius 
may reside at a large distance from the galaxy, due to
projection. The radial velocity data could be used to give 
a partial answer to distinguish halo PN from body PN. Multi-object 
spectroscopy techniques are required to obtain spectra
of the requisite numbers of PN to distinguish a trend and to
sample the line of sight abundance spread, excluding halo
objects. However given that most of the PN are projected against 
a strong stellar continuum, then multi-slit rather than multi-fibre
instruments are required to provide accurate background subtraction
of the spectra. Coherent fibre bundles, one for each PN, 
could alternatively be employed to allow effective 2-D background 
subtraction. However single fibres are adequate for
radial velocity work based on the brightest line of 5007\AA.
Since the orientation of a slit must be kept 
fixed to provide a good background subtraction and to allow several PN 
to be observed per slit, then the use of an Atmospheric Dispersion 
Corrector is highly advantageous to ensure spectrophotometry over the
requisite large wavelength range ($\sim$3700 - 6800\AA\ for He, N, O,
Ne and S abundance determinations). When considering observation
of  PN in galaxies more distant
than NGC~5128, the issue of background subtraction will become more
crucial as larger variations in galaxy continuum will be included
in the slit or aperture. Spectrophotometry of extra-galactic PN 
is a field where multi-object techniques on 8-10m telescopes will
bring a rich harvest of data to bear on the history of chemical
enrichment in galaxies of all types. 

\section{Conclusions}
The first spectra of planetary nebulae in the nearby early-type galaxy 
NGC~5128 have been presented. The spectra of five PN from the
catalogue of Hui et al. (\cite{huib}) have been
observed over an observed emission line brightness range of a 
factor 8 and galactocentric radius range from 7 to 18 kpc. The spectra show 
characteristic high ionization emission lines
similar to Galactic PN and confirming that the brightest PN in a galaxy
are entirely typical. The mean [O/H] of the five PN, determined
by empirical methods and modelling, is $-$0.5 with a spread of 
0.3dex. This low metallicity contrasts with that of the assumed metal 
rich stellar population of NGC~5128.

\begin{acknowledgements}
We would like to thank M. Richer for stimulating comments on 
the subject of probing galactic abundances from the PN
population.
\end{acknowledgements}

\noindent
{\bf Note added in proof} \\
Harris et al. (AJ 116, 2866, 1998 and AJ 117, 855, 1999) 
have obtained HST photometry of a globular cluster and the 
field halo stars in NGC~5128, situated at a distance of $\sim$21 kpc 
from the galaxy centre. For the halo stars they estimate a mean 
metallicity of $<[Fe/H]>$ = $-$0.4, but with a broad range. 
Although the PN observed in this paper were at smaller 
galactocentric distances, there is interesting agreement
between the metallicity of the PN from the [O/H] determinations 
presented here and those for
the Red Giant Branch stars observed by Harris et al.

\end{document}